\documentclass[12pt]{article}

\begin{document}
\renewcommand{\thefootnote}{\fnsymbol{footnote}}

\pagestyle{empty}

\begin{flushright}
hep-ph/9705428\\
May 1997\\[15mm]
\end{flushright}

\begin{center}
{\Large {\bf Two Remarks Concerning the Signs of the Light Quark
Masses}}\\[15mm] 
Henning Genz\\[2mm]
Special Research Centre for the Subatomic Structure of Matter\\[2mm]
The University of Adelaide\\[2mm]
South Australia 5005\\[15mm] 
{\Large {\bf Abstract}}\\
\end{center}
Partial conservation of the {\it nine} axial currents $A^\beta_\mu$ for
$\beta=0,1,...,8$ applied to the vacuum expectation values of the
equal-time commutators of the nine axial charges $ Q_A^\alpha(x_0) $ for
$\alpha=0,1,...,8$ with the corresponding axial current divergences
$\partial^\mu A_\mu^\beta $ implies that the non-vanishing current quark
masses $m_u,m_d,m_s$ of the light quarks $u, d, s$ have the same
sign. Under the more realistic assumption of partial conservation of only
the eight axial currents of $SU(3)\otimes SU(3)$, i.e. $\alpha=1,...,8$ and
$\beta=1,...,8$ in the above, inequalities for the quark masses
follow. They are trivially fulfilled if the three light quark masses have
the same sign and imply, for non-vanishing quark masses $m_u,m_d$, $m_s$,
that at least two of these masses have the same sign as their sum
$m_u+m_d+m_s$. If one of the three quark masses vanishes, one of the other
two might also. If both do not, they have the same sign (the same, of
course, as their sum). Our assumptions include the standard {\it vector}
$SU(3)$ symmetry of the vacuum.\\[5mm]

\begin{center}
(To appear in the Australian Journal of Physics)
\end{center}

\newpage
\pagestyle{plain}

\section{Introduction and conclusions}\label{res:section1}

Since quarks are confined, common sense arguments that all masses must have
the same (positive) sign cannot be applied to them. The present paper
derives restrictions on the signs of the light current quark masses from
the principles of quantum field theory, the standard model and Goldstone
chiral symmetry breaking.

Let us more precisely and maybe unrealistically first assume that in
computing the vacuum expectation values (VEV) $< [Q_A^\alpha (x_0),
\partial^\mu A_\mu^\beta (x)]>_0$ of the $\sigma$ commutators of the {\it
nine} axial charges $Q_A^\alpha$ with the {\it nine} axial current
divergencies $\partial^\mu A_\mu^\beta$ we may use partial conservation of
these currents. Thus for the present purpose the divergencies of the 16
vector and axial vector currents of $SU(3)\otimes SU(3)$ {\it and} of the
$SU(3)$-neutral ninth axial current $A_\mu^0$ may be computed from the
equation of motion of the standard model, or equivalently, from eq.(8)
together with the definition of $u(x)$ in eq.(3). The eight {\it vector}
charges $Q_V^a(x_0)$ for $a=1,...,8$ furthermore (almost) annihilate the
vacuum such that the VEVs of the vector $\sigma$-terms may be neglected
compared to the VEVs of the axial $\sigma$-terms.

This feature of chiral symmetry breaking is absolutely essential for our
conclusions. For example the replacement of the field $\psi_u$ of the
u-quark by $\gamma_5\psi_u$ flips the sign of $m_u$ and otherwise leaves
the QCD Lagrangian invariant. This is however not a counterexample to our
claims since under this transformation some of the vector currents such as
$\bar\psi_u\gamma_\mu\psi_d$ are taken into axial currents
($\bar\psi_u\gamma_\mu\gamma_5\psi_d$ in our example) such that, in the new
basis, the standard approximate $SU(3)$ symmetry of the vacuum, which we
assume, is replaced by a symmetry that is generated by a mixture of vector
and axial charges
\footnote{A much more detailed discussion of the relation
of the quark masses to the Goldstone nature of the vacuum, which can be
traced back to Dashen 1971, is contained in Leutwyler 1983. An early review
is Pagels 1975. I thank G. Ecker for asking a relevant question, an
anonymous referee for hints, and H. Leutwyler for sending me a copy of
Leutwyler 1983.}.

{}From the vacuum structure of the standard model and the fact that
$(-i)$-times the VEV of a $\sigma$-term of a charge $Q(x_0)=\int d^3 x
J_0(x)$ and the divergence $\partial^\mu J_\mu$ of the corresponding
hermitean current $J_\mu$ is non-negative,
$$-i<[Q, \partial^\mu J_\mu]>_0\geq 0,\eqno(1)$$ 
sum rules are derived that are fulfilled if and only if the non-vanishing
light quark current quark masses $m_u, m_d$ and $m_s$ have the same
(positive or negative) sign.

Restricting the above assumptions more realistically to the 16 vector and
axial vector currents of $SU(3)\otimes SU(3)$ and assuming that the sum
$m_u+m_d+m_s$ of the three quark masses is positive, we obtain the
relations
$$m_u+m_d\geq 0, \eqno(2a)$$
$$m_u+m_s\geq 0, \eqno(2b)$$
$$m_d+m_s\geq 0, \eqno(2c)$$
$$m_u+m_d+4m_s\geq 0\eqno(2d)$$
and
$$m_{u}m_{d}+m_{u}m_{s}+m_{d}m_{s}\geq 0\eqno(2e)$$ 
for the quark masses. They are trivially fulfilled if the masses of all
three quarks are non-negative and imply, for example, that at least two of
the three masses must be just that (i.e. non-negative). If one of the three
masses vanishes, a second one might. If not, the non-vanishing masses are
both positive. In the presumably academic case that the sum of the three
masses is negative, the less-or-equal sign in eqs.(2a)-(2d) are inverted,
whereas eq.(2e) remains unaltered. The conclusions remain very much the
same except that "non-negative" is replaced by "non-positive" and
"positive" by "negative". The case of a vanishing sum of the quark masses
is not discussed here.

\section{Reminders}

It is the purpose of the present section to remind the reader of the rather
old-fashioned methods, going back to Gell-Mann, Oakes, and Renner 1968, of
chiral symmetry breaking we use. The light quark mass term (averaging over
colour is always understood)
$$u(x)= m{_u}\bar\psi_u(x)\psi_u(x) +
m_d\bar\psi_d(x)\psi_d(x)+m_s\bar\psi_s(x)\psi_s(x)\eqno(3)$$ 
of the hamiltonian density of the standard model {\it and} chiral symmetry
breaking (see Leutwyler 1994 for a discussion of this) is written in the
form
$$u(x)=c_0u_0(x)+c_3u_3(x)+c_8u_8(x),\eqno(4)$$
where the definitions
$$c_0=(m_u+m_d+m_s)/\sqrt6, \eqno(5a) $$
$$c_3=(m_u-m_d)/2, \eqno(5b)$$
$$c_8=(m_u+m_d-2m_s)/(2\sqrt3) \eqno(5c)$$
together with
$$u_\alpha=\bar\psi (x)\lambda_{\alpha}\psi(x) \eqno(5d)$$
for $\alpha=0,1,...,8$ have been made. In the above, $\psi$ stands for the
three quark fields $\psi_u,\psi_d,\psi_s$ and the $\lambda_\alpha$ for
$\alpha=0,1,...,8$ are the well-known $3\times3$ Gell-Mann Matrices acting
on the components $\psi_u,\psi_d,\psi_s$ of $\psi$.

The complete Hamiltonian of the standard model can now be written as
$$H(x)=H_0(x)+u(x), \eqno(6)$$
where only $u(x)$ breaks $SU(3)\otimes SU(3)$ chiral symmetry. Namely,
defining the $SU(3)\otimes SU(3)$ vector and axial vector currents for
$a=1,...,8$ by $V_{\mu}^a=1/2\bar\psi\gamma_{\mu}\lambda_a\psi$ and
$A_{\mu}^a=1/2\bar\psi\gamma_5\gamma_{\mu}\lambda_a\psi$, respectively, the
corresponding charges $Q_V^a(x_0)=\int d^3xV_0^a(x)$ and $Q_A^a(x_0)=\int
d^3xA_0^a(x)$ commute with $H_0(x)$,
$$[H_0(x),Q(x_0)]=0, \eqno(7) $$
such that 
$$i[u(x),Q(x_0)]=\partial^{\mu}J_{\mu}(x) \eqno(8)$$
for $J_{\mu}$ any one of these currents and $Q(x_0)$ the corresponding
charge.

We remind the reader that eq.(8) can be derived from the equation of motion
of the standard model. More generally, eq.(8) follows from eqs.(6) and (7)
by use of the equal-time commutator (summation over $k=1,2,3$ is
understood)
$$i[H(x),J_0(y)]_{x_0=y_0=0}=\partial^\mu J_\mu(x)\delta^{(3)} (\vec x-\vec
y) +J_k(x)
\partial/\partial x_k\delta^{(3)} (\vec x-\vec y), \eqno(9)$$
which expresses (Genz and Katz 1971) the transformation properties of
$J_0(y)$ under time translations by $H=\int d^3x H(x)$ and boosts by
$M_{0k}=-\int d^3x x_k H(0,\vec x)$. 

Namely, integrating eq.(9) over $d^3y$ due to eq.(7) only $u(x)$ survives
in the equal-time commutator of $H(x)$ with the charge $Q(x_0)$. Computing
the equal-time commutators of the charges $Q_V^a$ and $Q_A^\alpha$ (with
$Q_A^0$ included) with the scalar densities $u_\alpha(x)$, which we have
already defined, and the pseudoscalar ones $v_\alpha=\bar\psi
(x)\gamma_5\lambda_{\alpha}\psi(x)$ one finds for latin indices between $1$
and $8$ and greek indices between $0$ and $8$ the equal-time commutation
relations (summation over double indices being understood; $f$ and $d$ are
the well-known Clebsch-Gordan coefficients of $SU(3)$)
$$[Q_V^a,u_\beta]=if_{a\beta c}u_c,\eqno(10a)$$
$$[Q_V^a,v_\beta]=if_{a\beta c}v_c,\eqno(10b)$$
$$[Q^\alpha_A,u_\beta]=id_{\alpha\beta\gamma}v_\gamma\eqno(10c)$$
and
$$[Q^\alpha_A,v_\beta]=-id_{\alpha\beta\gamma}u_\gamma.\eqno(10d)$$
For $\alpha\not=0$ these are the defining relations of the
$(3,\bar3)\oplus(\bar3,3)$ representation of $SU(3)\otimes SU(3)$.

As an immediate consequence, currents for which eq.(7) holds are conserved
in the limit of vanishing quark masses. Except for electromagnetic
anomalies, which are not supposed to contribute to the VEV of the
$\sigma$-commutators, Goldstone chiral symmetry breaking implies precisely
this for the 16 currents of $SU(3)\otimes SU(3)$. The ninth $SU(3)$-neutral
axial current $A_\mu^0$ is however not assumed to have a vanishing
divergence in this limit. Therefore we have stressed that our results which
involve the divergence of this particular current presumably are of
academic interest only.

The current divergencies turn out to be (summation once again being
understood)
$$\partial^\mu V_\mu^a=(c_3 f_{a3c}+c_8f_{a8c})u_c\eqno(11a)$$
and
$$\partial^\mu A_\mu^\alpha=( c_0 d_{\alpha0\gamma}+c_3 d_{\alpha3\gamma}
+c_8d_{\alpha8\gamma})v_\gamma.\eqno(11b)$$

It is now an easy exercise to compute the $\sigma$-commutators in terms of
the $u_\alpha$ and $v_\alpha$. The results for $u_\alpha$ imply the
reconstruction theorem of the chiral $SU(3)\otimes SU(3)$ symmetry breaking
Hamiltonian density which reads (Genz and Cornwell 1973)
$$u(x)=iC_R/2([Q_V^a,\partial^\mu V_\mu^a]+ [Q^a_A,\partial^\mu
A_\mu^a]),\eqno(12)$$ 
where $C_R=3/16$. Summation over $a=1,...,8$ is understood. For the
group-theoretically inclined reader we note that eq.(12) with $C_R>0$
follows for any $u$ that belongs to an irreducible representation of
$SU(3)\otimes SU(3)$ except the unit representation. The reason is that the
Casimir operator $1/2(Q_V^aQ_V^a+Q_A^aQ_A^a)$ of $SU(3)\otimes SU(3)$
acting on $u$ multiplies it with its (positive) Eigenvalue $C_R^{-1}$.

Thus, from eq.(1), we see that the VEV of $-u$ is non-negative (Genz and
Cornwell 1973)
$$<-u(x)>_0\geq 0,\eqno(13)$$
for any $u$ that belongs to an irreducible representation of $SU(3)\otimes
SU(3)$. If the equal-sign obtains, the divergences of all partially
conserved currents vanish.

Following Rausch 1982 we will exploit the VEV $<[Q_A^\alpha, \partial^\mu
A_\mu^\beta]>_0$ of the axial $\sigma$-terms and start with the remark that
nothing additional could be gained by also considering the VEV of the
others, i.e. $<[Q_V^a, \partial^\mu V_\mu^{b}]>_0$, $<[Q_V^a, \partial^\mu
A_\mu^\beta]>_0$ and $<[Q_A^\alpha, \partial^\mu V_\mu^{b}]>_0$, since
these either vanish trivially due to current or parity conservation or
since in the approximation in which we work the vector charges $Q_V^a$
($a=1,...,8$) annihilate the vacuum. This also implies that all scalar
densities $u_\alpha$ except $u_0$ have vanishing VEVs since they can be
represented as linear combinations of commutators of the type $[Q_V^a,u_b]$
such as e. g. $u_3=-i[Q_V^1,u_2]$ and
$u_8=i\sqrt3[Q_V^1,u_2]-i2/\sqrt3[Q_V^4,u_5]$. The vanishing of the VEVs of
$u_1,u_2,u_4,u_5,u_6$ and $u_7$ is also implied by the even stronger
argument that - except for special values of $c_3$ and $c_8$ - they are
proportional to a vector current divergence the VEV of which vanishes due
to translation invariance. In any case, we may conclude
$$<u_\alpha>_0=\delta_{0\alpha}<u_0>_0.\eqno(14)$$

Returning to the VEV $-i<[Q_A^\alpha, \partial^\mu A_\mu^\beta]>_0$ we will
have to distinguish between the $8\times8$ matrix for $\alpha$ and $\beta$
between 1 and 8 and the $9\times9$ matrix for the indices between 0 and
8. The $9\times9$ matrix can be written under our assumptions as
$$\sigma_{\alpha\beta}\equiv-i<[Q_A^\alpha, \partial^\mu A_\mu^\beta]>_0=$$
$$=\sqrt{2/3}<-c_0 u_0>_0(\delta_{\alpha\beta}+(c_3/c_0)d_{3\alpha\beta}
+(c_8/c_0)d_{8\alpha\beta}),\eqno(15)$$ 
which incidentally also shows that the matrix $\sigma_{\alpha\beta}$ is
symmetric under exchange of $\alpha$ and $\beta$. Making the definitions
$$G=2/3<u>_0=2/3<c_0 u_0>_0, \eqno(16a) $$
$$A=c_8/c_0\eqno(16b)$$
and
$$B=c_3/c_0\eqno(16c)$$ 
we can write the non-vanishing elements of the symmetric and real matrix
$\sigma$ as
$$\sigma_{00}=-G,\eqno(17a)$$
$$\sigma_{08} = \sigma_{80}=-GA,\eqno(17b)$$
$$\sigma_{03}=\sigma_{30}=-GB,\eqno(17c)$$
$$\sigma_{38}=\sigma_{83}=-GB/\sqrt2,\eqno(17d)$$
$$\sigma_{11}=\sigma_{22}=\sigma_{33}=-G(1+A/\sqrt2),\eqno(17e)$$
$$\sigma_{44}=\sigma_{55}=-G(1+B\sqrt{3/2}/2-A/(2\sqrt2)),\eqno(17f)$$
$$\sigma_{66}=\sigma_{77}=-G(1-B\sqrt{3/2}/2-A/(2\sqrt2))\eqno(17g)$$
and
$$\sigma_{88}=-G(1-A/\sqrt2).\eqno(17h)$$
Effectively, we therefore are dealing with a $4\times4$ or $5\times5$
matrix rather than an $8\times8$ or $9\times9$ matrix, respectively. It
should be noted that the fact that $\sigma{_{38}}$ does not vanish
immediately implies mixing of states with $\pi^0$ and $\eta$ quantum
numbers.

\section{The derivation}

{}From eq.(1) and the definition in eq.(15) it follows that for $z_m$ with
$m=0,8,3,4,6$ any five real numbers we have
$$z_mz_n\sigma_{mn}\geq0,\eqno(18)$$
where summation over $m$ and $n$ is understood. To exploit this condition
of positivity of the real symmetric $4\times4$ or $5\times5$ matrix
$\sigma$, we may apply (Rausch 1982) the Hausdorff criterion which states
that a $N\times N$ matrix $M_{ik}$ is positive if and only if for every $k$
between 1 and $N$

$$det\left(\matrix{ M_{11} &...& M_{1k}\cr ...&...&...\cr M_{k1}&...&
M_{kk}}\right)\geq0.$$

We start by exploiting partial conservation of the eight axial currents of
$SU(3)\otimes SU(3)$. From $\sigma_{33} + \sigma_{44}+ \sigma_{66}$ being
non-negative we obtain $-G\geq 0$, i.e. under more restrictive assumptions
once again eq.(13). It is gratifying that this result is in agreement with
$\sigma_{00}$ in eq.(17a) being non-negative. From $\sigma_{33}\geq0$ we
obtain the further result
$$A\geq -\sqrt2.\eqno(19a)$$
That the $2\times2$ submatrix $\sigma_{mn}$ for $m=3,8$ and $n=3,8$ is
non-negative yields
$$2\geq A^2+B^2\eqno(19b)$$
and from considering $\sigma_{44}$ and $\sigma_{55}$ separately we get
$$B\geq a/\sqrt3-2\sqrt{2/3}\eqno(19c)$$
and
$$2\sqrt{2/3}-A/\sqrt3\geq B.\eqno(19d)$$
Taken together, these relations are evidently somewhat redundant.

Replacing $Ac_0, Bc_0$ and $c_0$ by the corresponding linear combinations
of the quark masses and assuming that their sum is positive,
$$m_u+m_d+m_s>0,\eqno(20)$$
we obtain eqs.(2) as restrictions on these masses themselves. The changes
that occur if $m_u+m_d+m_s<0$ and the corresponding conclusions have
already be stated in connection with eqs.(2).

Finally, if our assumptions are extended to also cover the ninth axial
current $A^0_\mu$, we may apply the Hurwitz-criterion to the full
$5\times5$ matrix $\sigma$. It suffices to consider the cases $k=0,8,3$
with the results
$$-G\geq 0,\eqno(21a)$$
$$1-A/\sqrt2-A^2\geq 0\eqno(21b)$$
and
$$(\sqrt2a-1)[3/2B^2-(1+A/\sqrt2)^2]\geq 0.\eqno(21c)$$

As has already been said, eq.(21a) already follows from our previous
results. One easily sees that eq.(21b) is equivalent to
$$1/\sqrt2\geq A\geq-\sqrt2\eqno(22a)$$
such that from eq.(20c)
$$(1+a/\sqrt2)^2\geq3B^2\eqno(22b)$$
or equivalently
$$2+A^2+2\sqrt2 A\geq3B^2. \eqno(22c)$$
The relation $A\geq-\sqrt2$ has already been obtained in the above. Under
the assumption that the sum of the three quark masses is positive, the
additional relations are equivalent to
$$m_s\geq0\eqno(23a)$$
and
$$m_um_d\geq0. \eqno(23b)$$
Since we know that $m_u$ and $m_d$ cannot both be negative, they must be
positive (or at least one of them must be zero). If the sum of the quark
masses is negative, the greater-or-equal sign in eq.(23a) is inverted
whereas eq.(23b) remains unaltered. It then follows that that none of the
three quark masses can be positive.

Our conclusions have already been noted in connection with eq.(2). Under
the assumption that none of the three quark masses vanishes, they can be
stated as follows:
\begin{enumerate}
\item Partial conservation of the nine axial currents applied to the
computation of the VEV of the $\sigma$-terms in the standard model implies
that all three quark masses $m_u,m_d$ and $m_s$ have the same (positive or
negative) sign.
\item From partial conservation of only the eight axial currents of
$SU(3)\otimes SU(3)$ the inequalities in eqs.(2) follow. They imply for
example that at least two of the three quark masses have the same sign as
the sum $m_u+m_d+m_s$ of all of them.
\end{enumerate}
The reader is reminded that our assumptions include the standard {\it
vector} $SU(3)$ symmetry of the vacuum.

\section*{Acknowledgments}

I thank H. Leutwyler for encouragement to comment on the signs of the light
quark masses and R. Urech for a discussion of Genz and Cornwell 1973 and of
Rausch 1982 and for very valuable technical assistance. This work began in
1982 as a collaboration with my diploma-student H. Rausch. It was finished
in 1996 during a sabbatical leave from the Institut f\"ur Theoretische
Teilchenphysik der Universit\"at Karlsruhe, Germany. Thanks are due the
colleagues at the Special Research Centre for the Subatomic Structure of
Matter of the University of Adelaide for valuable discussions and
hospitality. A grant of the Volkswagen--Stiftung has made the visit possible.

\section*{References}
Dashen, R. (1971). Phys. Rev. {\bf D3}, 1879\newline\\
Gell-Mann, M., Oakes, R. J., and Renner, B. (1968). Phys. Rev. {\bf175},
2195\newline\\ 
Genz, H., and Cornwell, D. T. (1973). Fortschritte der Physik {\bf21},
559\newline\\
Genz, H., and Katz, J. (1971). Phys. Rev. {\bf D3}, 1860\newline\\
Leutwyler, H. (1983). On the vacuum angle in QCD; CERN-Report TH.3739
(unpublished)\newline\\
Leutwyler, H. (1994). Masses of the Light Quarks; Preprint
BUTP-94/8\newline\\
Pagels, H. (1975). Phys. Rep. {\bf16C}, 219\newline\\
Rausch, H. (1982). Einschr\"ankungen an die Quarkmassen durch die
Positivit\"at der $\sigma$-Terme bei gebrochener Isospinsymmetrie;
Diplomarbeit, University of Karlsruhe (unpublished)

\end{document}